\documentclass{webofc}
\usepackage[varg]{txfonts}
\usepackage{hyperref}
\usepackage{url}
\hypersetup{colorlinks=true,citecolor=blue,urlcolor=blue,linkcolor=blue}

\begin{document}
\title{Advancing multimessenger approaches in heavy-ion collisions: Insights from electromagnetic probes}

\author{\firstname{Lipei} \lastname{Du}\inst{1,2}\fnsep\thanks{
Invited plenary talk at the 12th International Conference on Hard and Electromagnetic Probes of High-Energy Nuclear Collisions (Hard Probes 2024), Nagasaki, Japan, Sep 22--27, 2024.}
}

\institute{Department of Physics, University of California, Berkeley CA 94270
\and Nuclear Science Division, Lawrence Berkeley National Laboratory, Berkeley CA 94270}

\abstract{Electromagnetic (EM) probes, including photons and dileptons, do not interact strongly after their production in heavy-ion collisions, allowing them to carry undistorted information from their points of origin. This makes them powerful tools for studying early-stage equilibration and the thermodynamic properties of the quark-gluon plasma (QGP). In these proceedings, we highlight recent theoretical advancements in EM probes, focusing on their role in probing early-stage dynamics and extracting medium properties. We also discuss the emerging multimessenger approach, which combines hadronic and electromagnetic probes to achieve a more comprehensive understanding of the QGP.}
\maketitle
%

\section{Introduction}

High-energy nuclear collisions provide a unique laboratory for studying strongly interacting matter under extreme conditions, recreating a state of matter known as the quark-gluon plasma (QGP), believed to have existed in the early universe shortly after the Big Bang. Understanding the properties of the QGP, including its thermodynamic characteristics, transport properties, and underlying microscopic interactions, is a central goal of heavy-ion physics \cite{Arslandok:2023utm,Sorensen:2023zkk,Du:2024wjm}. To achieve this, a variety of experimental probes are used to extract information about different aspects of the system’s evolution. 

Soft hadronic observables characterize the bulk medium, jet quenching reveals interactions between energetic partons and the medium, and weak probes provide insights into initial-state effects and nuclear parton distributions. Joining these, electromagnetic (EM) probes---photons and dileptons---hold a special place due to not interacting strongly with the medium, allowing them to escape the medium unaltered and carry direct information from their production points. This makes them valuable tools for studying both the early and thermalized stages of the QGP. 

In these proceedings, we focus on the theoretical progress in understanding EM probes, with an emphasis on their thermal emissions and what they reveal about the QGP’s equilibration processes and thermal properties. For recent short reviews on further advancements, see Refs.~\cite{Paquet:2023vkq,Vujanovic:2024bby,Gale:2025ome}.

\section{Challenges of electromagnetic probes}

Despite their unique advantages, interpreting electromagnetic probes poses significant challenges. Their spectra and yields contain contributions from multiple sources emitted throughout the collision, making it difficult to isolate specific signals \cite{Geurts:2022xmk}. For instance, photons originate from prompt hard scatterings, jet-medium interactions, thermal radiation, and hadronic decays, among others, while dileptons arise from Drell-Yan processes, QGP radiation, and hadronic decays, to name a few. Disentangling these contributions is essential for extracting meaningful physics. Kinematic selections can help isolate specific sources: high-$p_T$ photons are dominated by prompt production, while low-mass dileptons probe in-medium vector meson modifications \cite{Sakai:2023fbu,Rapp:2024grb,Atchison:2024lmf,Zhou:2024yyo}, intermediate-mass dileptons are sensitive to thermal radiation \cite{Rapp:2014hha,HADES:2019auv,STAR:2024bpc}, and high-mass dileptons largely originate from hard partonic processes. These challenges highlight the need for precise theoretical modeling and advanced experimental techniques to fully exploit the potential of EM probes in studying the QGP.

\section{Electromagnetic probes for QGP characterization}

In this section, we focus on photons and dileptons from thermal radiation, which provide essential information on the thermal properties of the QCD medium.

\subsection{Thermal spectra calculations}

The calculation of thermal spectra begins with the thermal emission rate, which is directly related to the electromagnetic spectral function $\text{Im}\, \Pi_{\text{em}}(q; T, \mu_B)$. This spectral function is derived from the retarded photon self-energy $\Pi_{\text{em}}^{\mu\nu}$, a correlation function of electromagnetic currents that encapsulates all QCD corrections. Here, $q$ is the four-momentum of the emitted photon or dilepton, $T$ is the temperature, and $\mu_B$ is the baryon chemical potential of a thermal source. The spectral function can be calculated using two rigorous methods: perturbative QCD (pQCD) and lattice QCD. Each method has its strengths and limitations.

\begin{figure}[h]
    \centering
    \raisebox{0.4\height}{\includegraphics[width=0.45\linewidth]{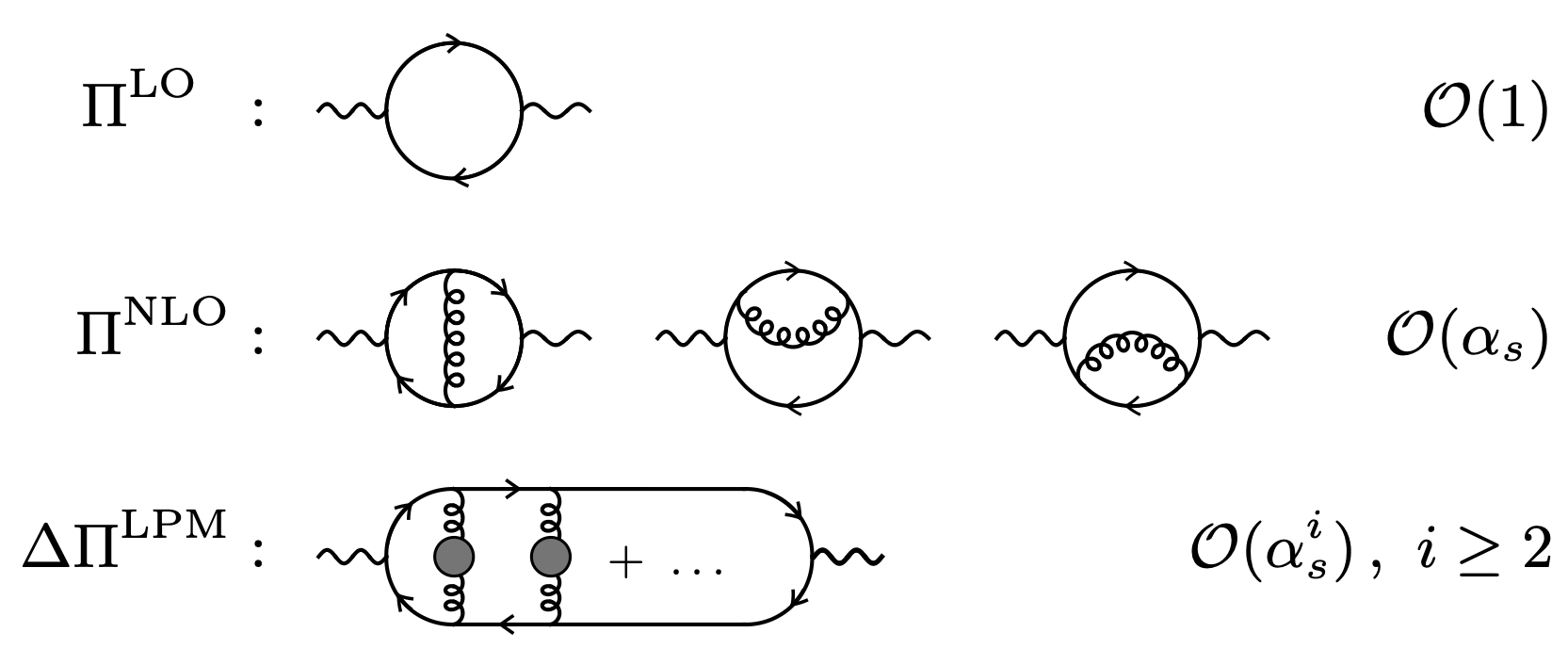}}\hspace{0.5cm}
    \includegraphics[width=0.45\linewidth]{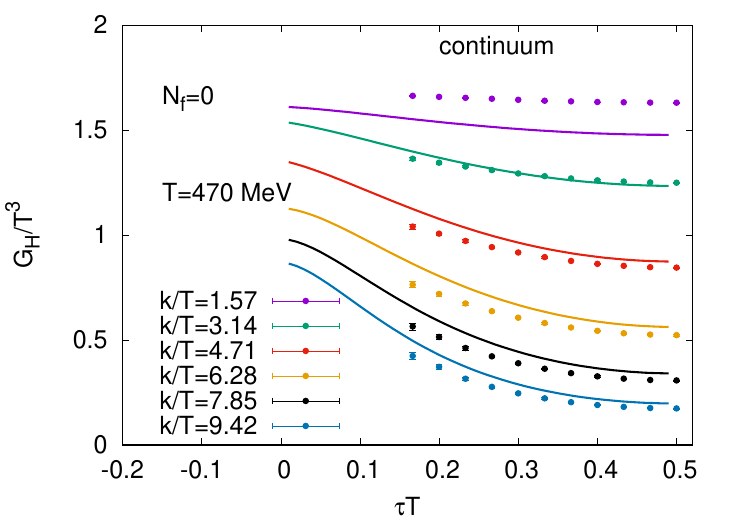}
    \caption{Left: Diagrams included in the evaluation of QCD corrections to the dilepton emission rate using perturbative QCD methods, as calculated in Refs.~\cite{Ghisoiu:2014mha,Churchill:2023zkk,Churchill:2023vpt}. Right: Euclidean correlator computed from the perturbative spectral function (solid lines) using the method shown on the left, compared to lattice QCD measurements (points) for several momenta \cite{Ali:2024xae}.}
    \label{fig:spectral}
\end{figure}

Lattice QCD is fully non-perturbative and provides first-principles calculations, but is limited to imaginary time, making real-time spectral functions challenging to extract. Perturbative QCD allows for real-time calculations and has been extended to next-to-leading order (NLO) for both photons \cite{Ghiglieri:2013gia} and dileptons \cite{Laine:2013vma,Churchill:2023vpt,Churchill:2023zkk}. However, pQCD calculations may face convergence issues, particularly at low temperatures. The complexity of pQCD calculations is illustrated by the photon self-energy diagrams shown in the left panel of Fig.~\ref{fig:spectral}. These diagrams, which include the strict NLO part as well as Landau-Pomeranchuk-Migdal (LPM) resummation, highlight the intricate interplay between QCD interactions and electromagnetic emissions in the QGP. Recent advancements have led to good agreement between lattice QCD and pQCD calculations \cite{Ali:2024xae}. For example, state-of-the-art comparisons, as shown in the right panel of Fig.~\ref{fig:spectral}, demonstrate that pQCD calculations at NLO can reasonably reproduce lattice results for the spectral function.

In addition to equilibrium calculations, off-equilibrium corrections and magnetic field effects have been incorporated into emission rates. Off-equilibrium corrections account for viscous effects, which are essential for understanding hadronic observables. For a detailed discussion of these corrections, including their impact on anisotropic flows due to bulk and shear viscous effects, see a recent review \cite{Vujanovic:2024bby}. Magnetic fields, generated by the relativistic motion of charged particles, can also modify spectral functions and affect anisotropies in photon and dilepton emissions. For recent progress in this area, see, e.g., Refs. \cite{gao:2024,Sun:2023pil}.

\subsection{Probing QGP equilibration}\label{sec:equilibration}

In the early stages of a heavy-ion collision, the system is far from equilibrium, characterized by strong momentum anisotropies and rapid expansion. During this pre-equilibrium phase, the system begins with a gluon-dominated initial state, and partonic interactions drive the system toward local thermal and chemical equilibrium before transitioning to a hydrodynamic description. Kinetic transport approaches are essential tools for modeling this complex regime. These methods solve the Boltzmann equation for partons, accounting for scatterings and interactions that govern the system's evolution \cite{Massen:2024pnj,Coquet:2021lca,Garcia-Montero:2023lrd}. By tracking the distribution functions of quarks and gluons, these approaches can describe the pre-equilibrium EM radiation during this early phase. 

\begin{figure}[h]
    \centering
    \includegraphics[width=0.45\linewidth]{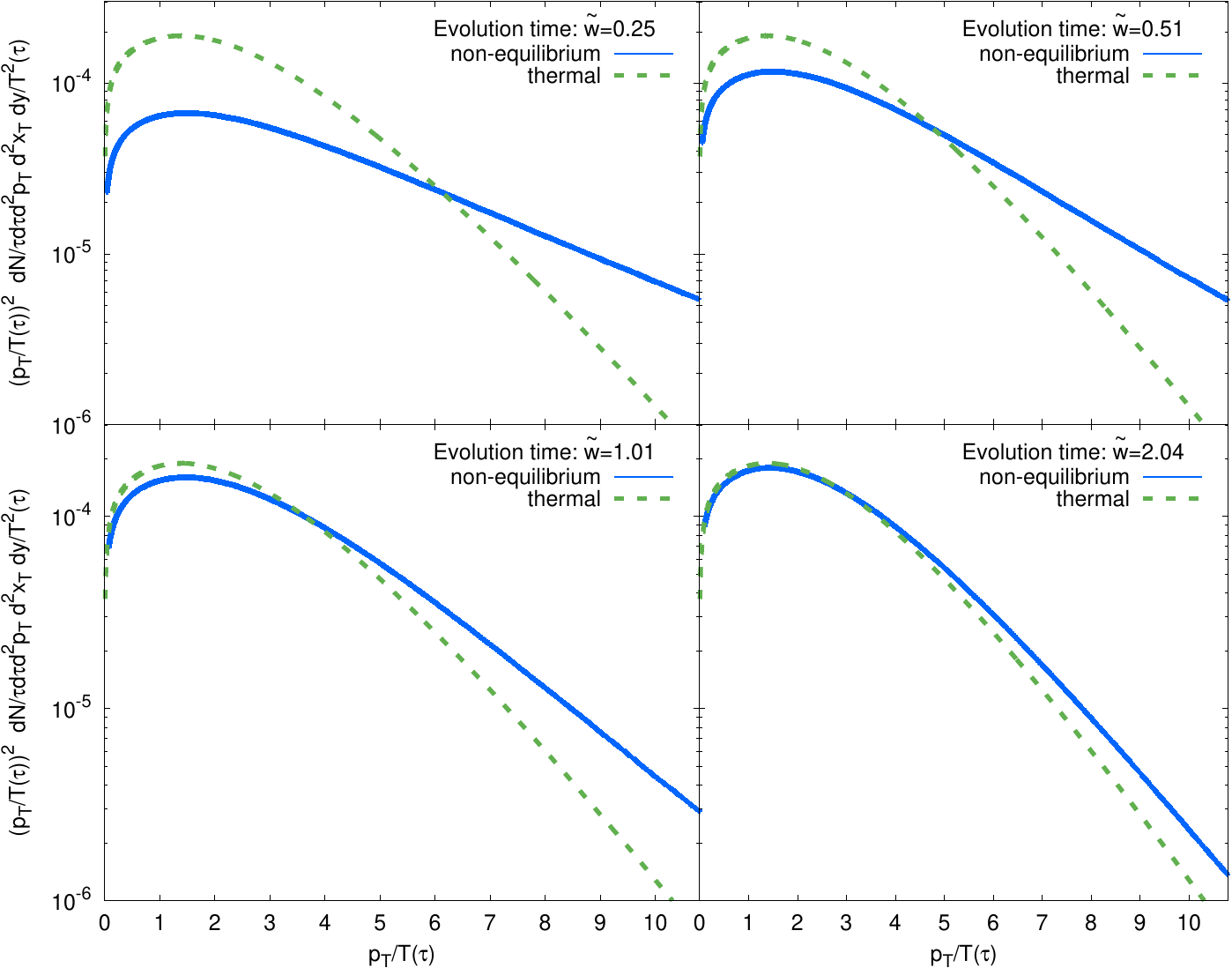}\hspace{0.5cm}
    \includegraphics[width=0.38\linewidth]{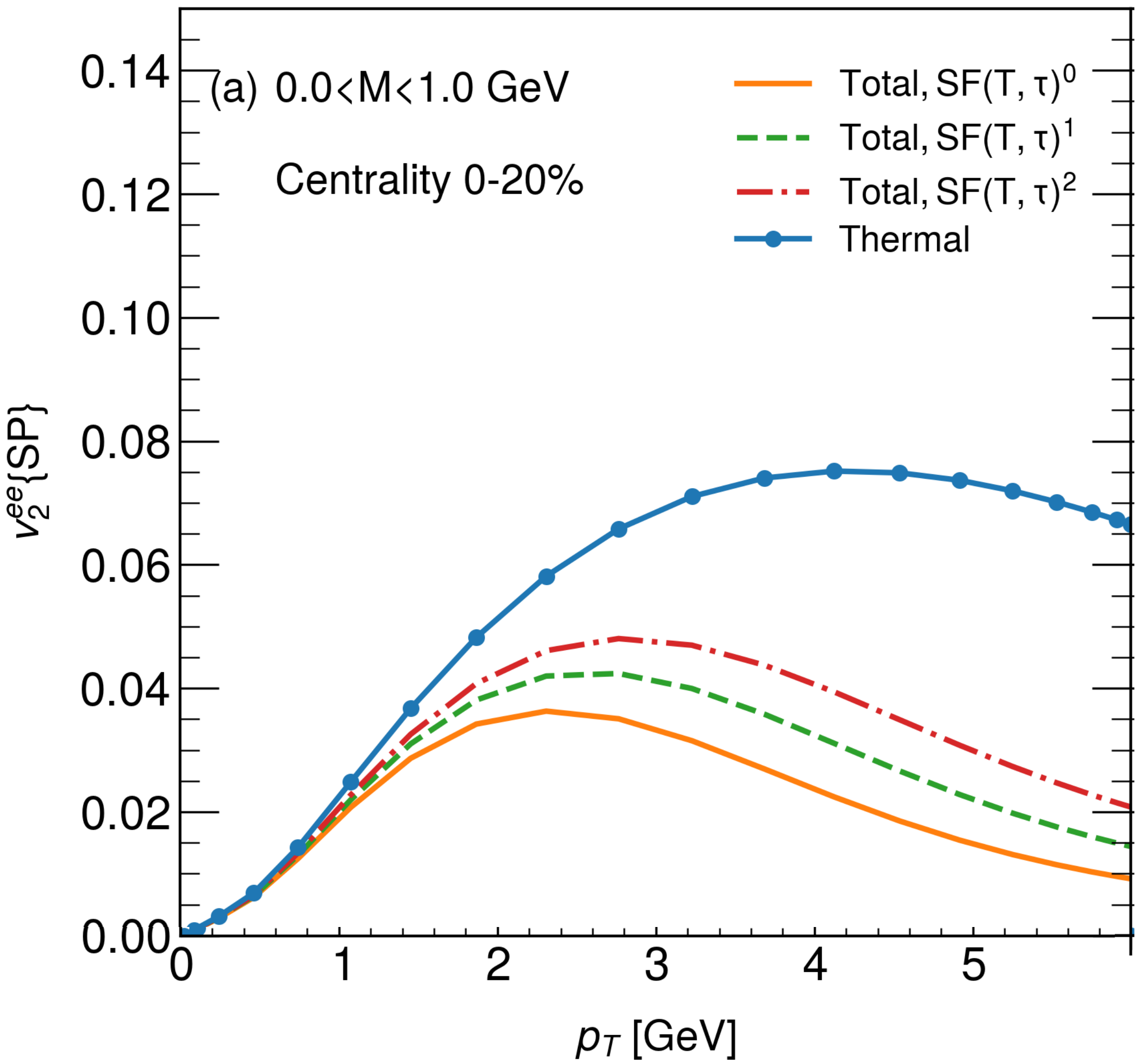}
    \caption{Left: The photon emission rate at leading order, showing non-equilibrium production (blue solid curves) and the thermal contribution (green dashed lines) \cite{Garcia-Montero:2023lrd}.
    Right: The thermal and total dilepton elliptic flow as a function of transverse momentum, with varying rates of chemical equilibrium. \cite{Wu:2024pba}}
    \label{fig:equilibration}
\end{figure}

Advances in kinetic theory and transport models have enabled more accurate calculations of pre-equilibrium EM radiation, bridging the gap between early-time dynamics and hydrodynamic evolution. These findings provide valuable insights into using EM emission to probe the dynamics of quark and gluon equilibration in high-energy collisions. In Ref.~\cite{Garcia-Montero:2023lrd}, kinetic approaches reveal that the non-equilibrium photon spectrum is significantly below the thermal spectrum at low transverse momentum and is much harder, with thermalization first achieved in the soft regime (left panel of Fig.~\ref{fig:equilibration}). Another study, employing a multistage hydrodynamic model with K\o{}MP\o{}ST as the pre-equilibrium dynamic model, investigates dilepton spectra and anisotropic flows in Pb-Pb collisions at $\sqrt{s_{NN}}=5.02$ TeV \cite{Wu:2024pba}. The results show that faster equilibration leads to larger yields and smaller $v_n(p_T)$, indicating that the combination of spectra and anisotropic flows is a powerful tool for probing equilibration (right panel of Fig.~\ref{fig:equilibration}). Similar insights are applicable to photons \cite{Gale:2021emg}.

Additionally, in Ref.~\cite{Coquet:2023wjk}, the angular distribution of dileptons with respect to the beam axis is explored as a probe of equilibration, particularly in the context of Drell-Yan processes and pre-equilibrium QGP. The quadrupole moment of the angular distribution is discussed, with Drell-Yan processes exhibiting a positive quadrupole moment due to preferential longitudinal emission, while pre-equilibrium QGP shows a negative quadrupole moment due to preferential transverse emission. The total quadrupole moment can change sign at varying invariant mass, depending on the dominance of pre-equilibrium QGP emission, which is influenced by the degree of equilibration.

\subsection{Electromagnetic probes as thermometers}

Electromagnetic radiation provides a direct link between observed spectra and the temperature of the emitting medium \cite{Rapp:2014hha,Shen:2013vja}. In the non-relativistic limit ($M \gg T$), the dilepton invariant mass ($M$) spectrum follows $(MT)^{3/2}e^{-M/T}$, while for photons with $p_T \gg T$, the transverse momentum ($p_T$) spectrum behaves as $e^{-p_T/T}$. These exponential dependencies offer a natural way to extract temperatures from experimental data. However, photon $p_T$ spectra are affected by blue shifts due to collective expansion, reflecting both thermal properties and bulk motion. In contrast, dilepton invariant mass spectra are largely insensitive to flow effects, making them a more direct probe of the system's temperature.

\begin{figure}[h]
    \centering
    \includegraphics[width=0.43\linewidth]{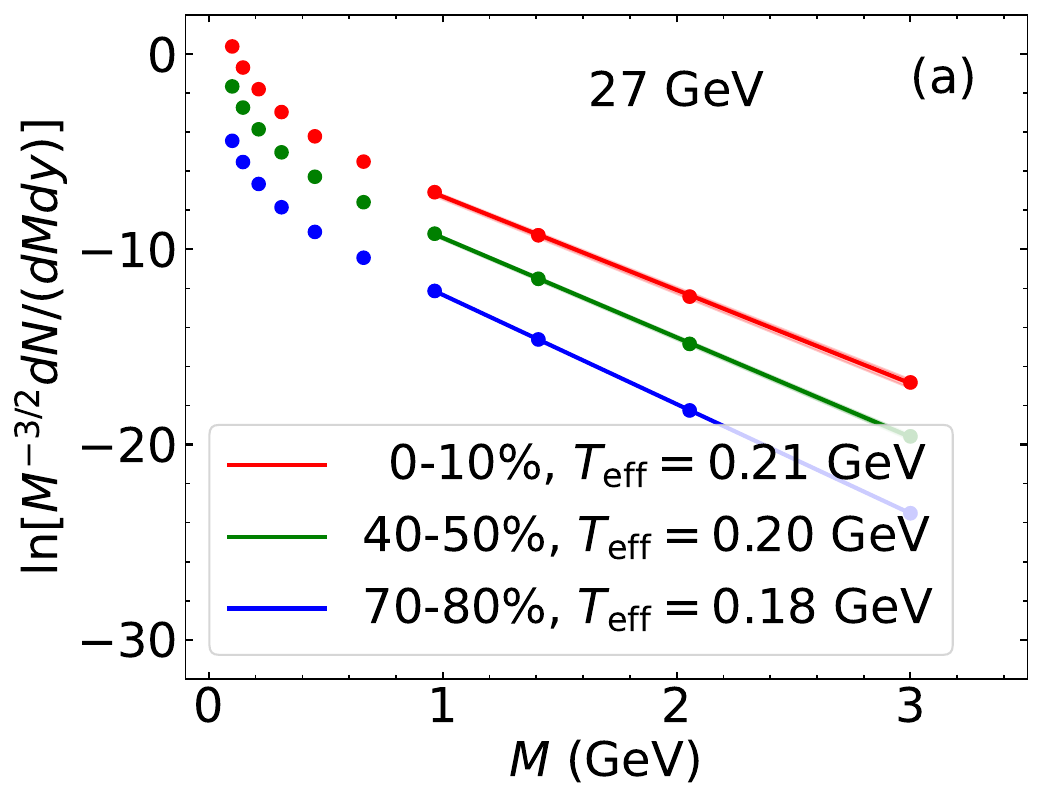}\hspace{1cm}
    \includegraphics[width=0.44\linewidth]{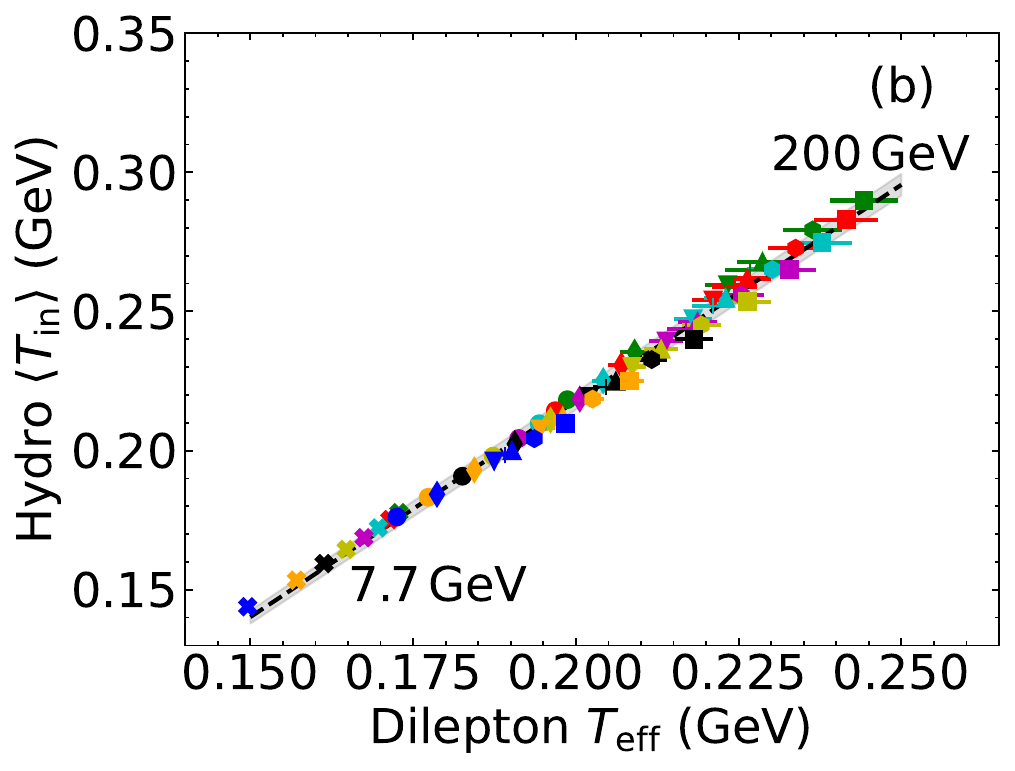}
    \caption{Left: Effective temperature $T_{\text{eff}}$ extracted from the dilepton spectrum for Au+Au collisions at $\sqrt{s_{NN}} = 27$ GeV. The straight lines represent linear fits in the mass range $1 \leq M \leq 3$ GeV, with the inverse slope providing $T_{\text{eff}}$ \cite{Churchill:2023vpt}. 
    Right: Correlation between the initial average hydrodynamic temperature $\langle T_{\text{in}} \rangle$ and the extracted $T_{\text{eff}}$ from dilepton spectra for energies from 7.7 to 200 GeV. Horizontal error bars reflect uncertainties from the fitting procedure shown in the left panel \cite{Churchill:2023zkk}.}
    \label{fig:thermometer}
\end{figure}

Despite the utility of these spectra for temperature extraction in experiments \cite{HADES:2019auv,STAR:2024bpc}, two major challenges persist. First, the approximate formulas apply for a static, homogeneous system, while the medium in heavy-ion collisions is highly dynamic and inhomogeneous. It remains unclear whether fitting experimental data with these formulas accurately characterizes the system, especially given current measurement uncertainties. Second, even if meaningful temperatures are extracted, they represent effective or averaged quantities over the entire space-time evolution, rather than well-defined thermodynamic properties. Providing a more physically meaningful interpretation for these temperatures is not straightforward.

Interpreting these temperatures requires advanced theoretical modeling, incorporating state-of-the-art descriptions of medium evolution \cite{Du:2022yok,Du:2023gnv,Du:2023efk,Massen:2024pnj} and electromagnetic emission rates to connect observed spectra with the underlying medium properties. Addressing these challenges, Refs.~\cite{Churchill:2023zkk,Churchill:2023vpt} investigate dilepton spectra for Au+Au collisions at $\sqrt{s_{NN}}$ from 7.7 to 200 GeV. First, they demonstrate that the fitting method works well when applied to model-calculated spectra in the mass range $1 \leq M \leq 3$ GeV [Fig.~\ref{fig:thermometer}(a)]. Second, they reveal a strong correlation between the extracted temperature from dilepton spectra and the average initial temperature  [Fig.~\ref{fig:thermometer}(b)]. This not only provides a more physically meaningful interpretation of the extracted temperature but also offers a way to measure the initial temperature of the highly dynamic QCD medium, which is immune to dynamical distortions. This represents a significant advancement in using electromagnetic probes as reliable thermometers of the QCD medium in heavy-ion collision studies.

\subsection{Multimessenger approaches}

The study of heavy-ion collisions has mainly relied on individual probes, such as soft hadronic observables and electromagnetic emissions, to extract key properties of the QGP. However, a multimessenger approach, which systematically combines different types of signals, has emerged as a powerful method to gain a more comprehensive understanding of the system’s evolution \cite{Gale:2021emg,Du:2024pbd}. Each class of observables captures distinct aspects of the collision: for instance, soft hadrons provide information about the system at freeze-out, while electromagnetic probes offer insights into the entire space-time evolution. By correlating these complementary signals, multimessenger studies enable a more robust extraction of fundamental QGP properties, reduce model dependencies, and offer new ways to disentangle overlapping physics effects. Additionally, studying multiple messengers within a consistent medium evolution framework, identifying their correlations, and testing these correlations in experiments provide a robust test of the underlying physics in the model.

\begin{figure}[h]
    \centering
    \includegraphics[width=0.45\linewidth]{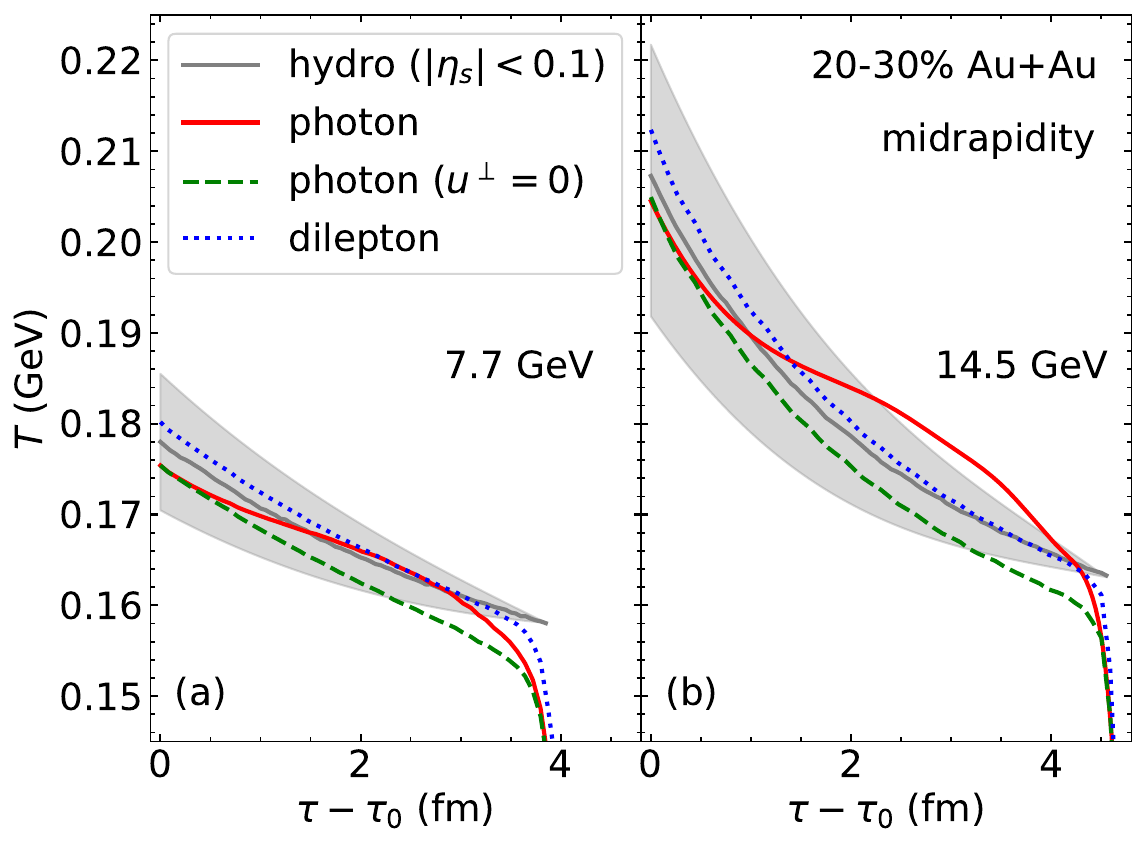}\hspace{0.5cm}
    \includegraphics[width=0.45\linewidth]{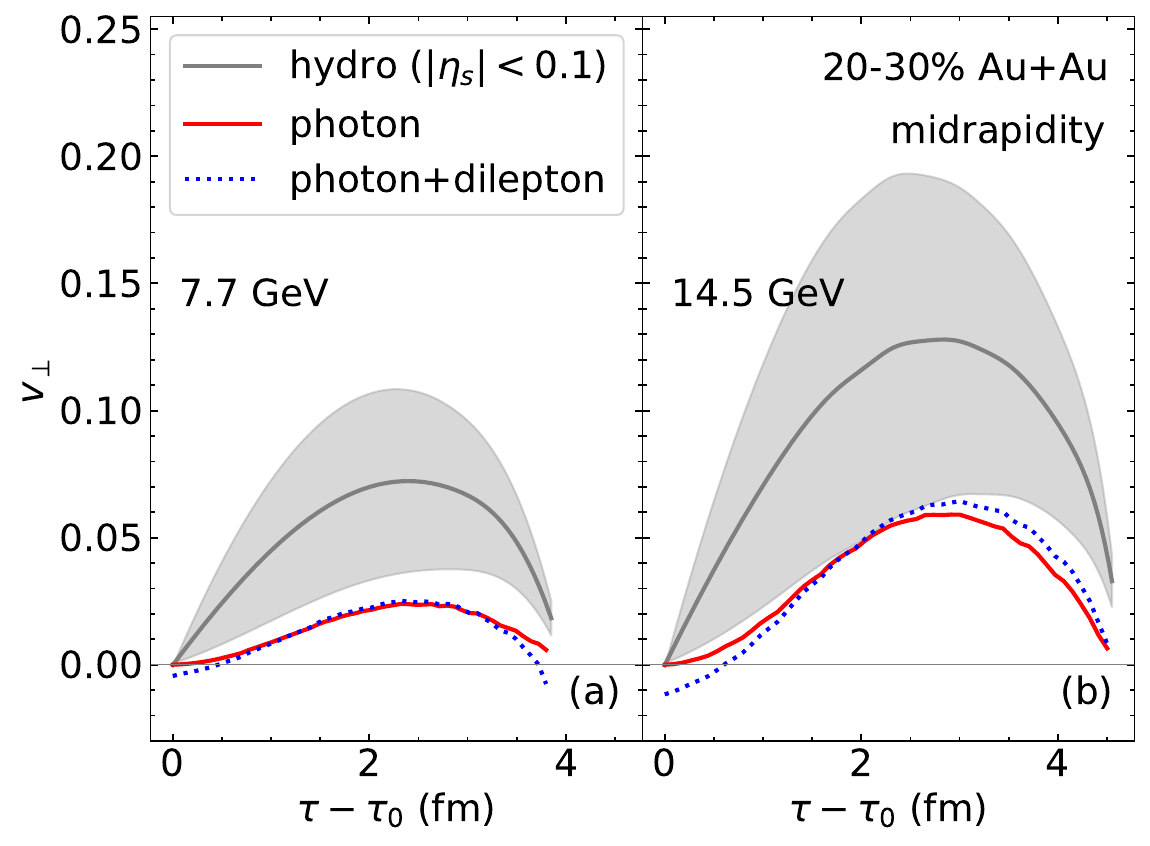}
    \caption{Left: Temperature evolution at midrapidity for 20–30\% Au+Au collisions at (a) 7.7 GeV and (b) 14.5 GeV. Gray curves show the weighted average and variation of hydrodynamic cell temperatures within $|\eta_s| < 0.1$. Temperatures from photon $p_T$ spectra are shown with (red solid) and without (green dashed) transverse flow, while blue dotted lines show temperatures from dilepton invariant mass spectra.  
    Right: Extracted radial flow. Red solid line combines photon temperatures with and without transverse flow; blue dotted line combines photon temperatures with transverse flow and dilepton temperatures \cite{Du:2024pbd}.
    }
    \label{fig:multi}
\end{figure}

The multimessenger approach is particularly valuable for studying collective dynamics, such as radial and anisotropic flow, where different probes experience varying levels of interaction with the medium, allowing for independent constraints on collectivity and equilibration. In Sec.~\ref{sec:equilibration}, we discussed studies that investigate the anisotropic flows of both soft hadrons and electromagnetic probes to constrain chemical equilibration in the early stages of evolution. Recently, efforts have been made to explore correlations between temperatures extracted from photon $p_T$ spectra, dilepton $M$ spectra, and hydrodynamic temperatures tuned by soft hadronic observables (left panels of Fig.~\ref{fig:multi}) \cite{Du:2024pbd}. A correlation has been demonstrated between temperatures extracted from photon spectra without Doppler shift (green dashed lines) and dilepton spectra (blue dotted lines). This correlation suggests that combining temperatures extracted from measurable photon spectra (which undergo Doppler shift, red solid lines) with those from dilepton measurements (which provide temperatures without Doppler shift) can offer valuable insights into the development of collective radial flow (right panels of Fig.~\ref{fig:multi}).

\section{Summary and outlook}

Recent advancements in both theory and experiment have significantly improved our understanding of electromagnetic probes in heavy-ion collisions. High-precision measurements of photon and dilepton spectra at RHIC and the LHC, combined with state-of-the-art dynamic models incorporating emission rates, have provided deeper insights into the QGP's chemical equilibration, temperature, and collective flow. These developments have refined our ability to extract key properties of the medium, shedding light on its thermal and non-equilibrium evolution.

Despite these advances, challenges remain in disentangling contributions from different stages of the collision. Direct photon spectra include prompt, pre-equilibrium, and thermal radiation components, requiring careful modeling to isolate medium-induced signals. Dilepton spectra, while valuable for temperature extraction, are affected by medium modifications of vector mesons and non-equilibrium effects, complicating their interpretation. Addressing these uncertainties requires further improvements in theoretical frameworks, particularly in incorporating pre-equilibrium dynamics, in-medium spectral modifications, and more precise treatments of electromagnetic emission rates. Recent progress in off-equilibrium corrections and the influence of strong magnetic fields has also highlighted the need for advanced modeling to account for these effects on emission rates.

Looking ahead, the future of heavy-ion physics will increasingly rely on multimessenger approaches, where electromagnetic probes are analyzed alongside hadronic observables to provide a more complete picture of QGP evolution. By combining photons, dileptons, and hadrons within a consistent medium evolution framework, we can identify correlations between these probes and test them experimentally, offering a robust validation of the underlying physics. The use of Bayesian analysis to systematically combine multimessenger data will further reduce model dependencies and provide tighter constraints on QGP properties, such as temperature, viscosity, and initial conditions \cite{JETSCAPE:2020mzn,JETSCAPE:2024cqe}.

The next-generation facilities and detector upgrades at RHIC and the LHC will significantly enhance precision measurements of electromagnetic probes across a broader kinematic range. Future experiments at GSI/FAIR, HIAF, and NICA will further explore the QGP under different temperature and density conditions, offering new opportunities to probe its fundamental properties. As theoretical modeling continues to evolve and experimental capabilities expand, electromagnetic probes will remain indispensable tools for advancing our knowledge of extreme QCD matter in both heavy-ion collisions and other astrophysical systems.

~\\
\textbf{Acknowledgments ---}
I would like to thank the organizers of the 12th International Conference on Hard and Electromagnetic Probes of High-Energy Nuclear Collisions for their kind invitation to present this work. I also thank Charles Gale, Greg Jackson, Peter Jacobs, Sangyong Jeon, Jean-François Paquet, Chun Shen, Gojko Vujanovic, and Xiang-Yu Wu for their helpful discussion. Some of the studies covered in these proceedings were partly supported by the
Natural Sciences and Engineering Research Council of Canada. I thank the collaborators and researchers whose work contributed to the results discussed here.

\bibliography{refs}

\begin{thebibliography}{35}

\bibitem{Arslandok:2023utm}
M.~Arslandok et~al., {Hot QCD White Paper} (2023), \texttt{2303.17254}.

\bibitem{Sorensen:2023zkk}
A.~Sorensen et~al., {Dense nuclear matter equation of state from heavy-ion
  collisions}, Prog. Part. Nucl. Phys. \textbf{134}, 104080 (2024),
  \texttt{2301.13253}. \doiwoc{10.1016/j.ppnp.2023.104080}

\bibitem{Du:2024wjm}
L.~Du, A.~Sorensen, M.~Stephanov, {The QCD phase diagram and Beam Energy Scan
  physics: a theory overview}, Int. J. Mod. Phys. E \textbf{33}, 2430008
  (2024), \texttt{2402.10183}. \doiwoc{10.1142/S021830132430008X}

\bibitem{Paquet:2023vkq}
J.F. Paquet, {Electromagnetic probes in heavy-ion collisions: progress and open
  questions}, PoS \textbf{HardProbes2023}, 009 (2024), \texttt{2307.09967}.
  \doiwoc{10.22323/1.438.0009}

\bibitem{Vujanovic:2024bby}
G.~Vujanovic, {Electromagnetic Probes of the Quantum Chromodynamical Plasma},
  in \emph{{14th International Conference on Nucleus Nucleus Collisions}}
  (2024), \texttt{2411.19868}

\bibitem{Gale:2025ome}
C.~Gale, {Electromagnetic Radiation from High-Energy Nuclear Collisions}
  (2025), \texttt{2502.13938}

\bibitem{Geurts:2022xmk}
F.~Geurts, R.A. Tripolt, {Electromagnetic probes: Theory and experiment}, Prog.
  Part. Nucl. Phys. \textbf{128}, 104004 (2023), \texttt{2210.01622}.
  \doiwoc{10.1016/j.ppnp.2022.104004}

\bibitem{Sakai:2023fbu}
A.~Sakai, M.~Harada, C.~Nonaka, C.~Sasaki, K.~Shigaki, S.~Yano, {Probing the
  QCD phase transition with chiral mixing in dilepton production} (2023),
  \texttt{2308.03305}.

\bibitem{Rapp:2024grb}
R.~Rapp, {Electric conductivity of QCD matter and dilepton spectra in heavy-ion
  collisions}, Phys. Rev. C \textbf{110}, 054909 (2024), \texttt{2406.14656}.
  \doiwoc{10.1103/PhysRevC.110.054909}

\bibitem{Atchison:2024lmf}
J.~Atchison, Y.~Han, F.~Geurts, {Electric conductivity of hot and dense nuclear
  matter}, Phys. Lett. B \textbf{858}, 139024 (2024), \texttt{2408.10176}.
  \doiwoc{10.1016/j.physletb.2024.139024}

\bibitem{Zhou:2024yyo}
W.H. Zhou, C.M. Ko, K.J. Sun, {Effects of chiral symmetry restoration on
  dilepton production in heavy ion collisions} (2024), \texttt{2412.18895}.

\bibitem{Rapp:2014hha}
R.~Rapp, H.~van Hees, {Thermal Dileptons as Fireball Thermometer and
  Chronometer}, Phys. Lett. B \textbf{753}, 586 (2016), \texttt{1411.4612}.
  \doiwoc{10.1016/j.physletb.2015.12.065}

\bibitem{HADES:2019auv}
J.~Adamczewski-Musch et~al. (HADES), {Probing dense baryon-rich matter with
  virtual photons}, Nature Phys. \textbf{15}, 1040 (2019).
  \doiwoc{10.1038/s41567-019-0583-8}

\bibitem{STAR:2024bpc}
{Temperature Measurement of Quark-Gluon Plasma at Different Stages} (2024),
  \texttt{2402.01998}.

\bibitem{Ghisoiu:2014mha}
I.~Ghisoiu, M.~Laine, {Interpolation of hard and soft dilepton rates}, JHEP
  \textbf{10}, 083 (2014), \texttt{1407.7955}. \doiwoc{10.1007/JHEP10(2014)083}

\bibitem{Churchill:2023zkk}
J.~Churchill, L.~Du, C.~Gale, G.~Jackson, S.~Jeon, {Virtual Photons Shed Light
  on the Early Temperature of Dense QCD Matter}, Phys. Rev. Lett. \textbf{132},
  172301 (2024), \texttt{2311.06951}. \doiwoc{10.1103/PhysRevLett.132.172301}

\bibitem{Churchill:2023vpt}
J.~Churchill, L.~Du, C.~Gale, G.~Jackson, S.~Jeon, {Dilepton production at
  next-to-leading order and intermediate invariant-mass observables}, Phys.
  Rev. C \textbf{109}, 044915 (2024), \texttt{2311.06675}.
  \doiwoc{10.1103/PhysRevC.109.044915}

\bibitem{Ali:2024xae}
S.~Ali, D.~Bala, A.~Francis, G.~Jackson, O.~Kaczmarek, J.~Turnwald, T.~Ueding,
  N.~Wink (HotQCD), {Lattice QCD estimates of thermal photon production from
  the QGP}, Phys. Rev. D \textbf{110}, 054518 (2024), \texttt{2403.11647}.
  \doiwoc{10.1103/PhysRevD.110.054518}

\bibitem{Ghiglieri:2013gia}
J.~Ghiglieri, J.~Hong, A.~Kurkela, E.~Lu, G.D. Moore, D.~Teaney,
  {Next-to-leading order thermal photon production in a weakly coupled
  quark-gluon plasma}, JHEP \textbf{05}, 010 (2013), \texttt{1302.5970}.
  \doiwoc{10.1007/JHEP05(2013)010}

\bibitem{Laine:2013vma}
M.~Laine, {NLO thermal dilepton rate at non-zero momentum}, JHEP \textbf{11},
  120 (2013), \texttt{1310.0164}. \doiwoc{10.1007/JHEP11(2013)120}

\bibitem{gao:2024}
H.~Gao, et~al. (2024), {In these proceedings.}

\bibitem{Sun:2023pil}
J.A. Sun, L.~Yan, {The effect of weak magnetic photon emission from quark-gluon
  plasma}, Phys. Lett. B \textbf{858}, 139046 (2024), \texttt{2302.07696}.
  \doiwoc{10.1016/j.physletb.2024.139046}

\bibitem{Massen:2024pnj}
O.~Massen, G.~Nijs, M.~Sas, W.~van~der Schee, R.~Snellings, {Effective
  temperatures of the QGP from thermal photon and dilepton production} (2024),
  \texttt{2412.09671}.

\bibitem{Coquet:2021lca}
M.~Coquet, X.~Du, J.Y. Ollitrault, S.~Schlichting, M.~Winn, {Intermediate mass
  dileptons as pre-equilibrium probes in heavy ion collisions}, Phys. Lett. B
  \textbf{821}, 136626 (2021), \texttt{2104.07622}.
  \doiwoc{10.1016/j.physletb.2021.136626}

\bibitem{Garcia-Montero:2023lrd}
O.~Garcia-Montero, A.~Mazeliauskas, P.~Plaschke, S.~Schlichting,
  {Pre-equilibrium photons from the early stages of heavy-ion collisions}, JHEP
  \textbf{03}, 053 (2024), \texttt{2308.09747}.
  \doiwoc{10.1007/JHEP03(2024)053}

\bibitem{Wu:2024pba}
X.Y. Wu, L.~Du, C.~Gale, S.~Jeon, {Probing the equilibration of the QCD matter
  created in heavy-ion collisions with dileptons}, Phys. Rev. C \textbf{110},
  054904 (2024), \texttt{2407.04156}. \doiwoc{10.1103/PhysRevC.110.054904}

\bibitem{Gale:2021emg}
C.~Gale, J.F. Paquet, B.~Schenke, C.~Shen, {Multimessenger heavy-ion collision
  physics}, Phys. Rev. C \textbf{105}, 014909 (2022), \texttt{2106.11216}.
  \doiwoc{10.1103/PhysRevC.105.014909}

\bibitem{Coquet:2023wjk}
M.~Coquet, M.~Winn, X.~Du, J.Y. Ollitrault, S.~Schlichting, {Dilepton
  Polarization as a Signature of Plasma Anisotropy}, Phys. Rev. Lett.
  \textbf{132}, 232301 (2024), \texttt{2309.00555}.
  \doiwoc{10.1103/PhysRevLett.132.232301}

\bibitem{Shen:2013vja}
C.~Shen, U.W. Heinz, J.F. Paquet, C.~Gale, {Thermal photons as a quark-gluon
  plasma thermometer reexamined}, Phys. Rev. C \textbf{89}, 044910 (2014),
  \texttt{1308.2440}. \doiwoc{10.1103/PhysRevC.89.044910}

\bibitem{Du:2022yok}
L.~Du, C.~Shen, S.~Jeon, C.~Gale, {Probing initial baryon stopping and
  equation~of state with rapidity-dependent directed flow of identified
  particles}, Phys. Rev. C \textbf{108}, L041901 (2023), \texttt{2211.16408}.
  \doiwoc{10.1103/PhysRevC.108.L041901}

\bibitem{Du:2023gnv}
L.~Du, H.~Gao, S.~Jeon, C.~Gale, {Rapidity scan with multistage hydrodynamic
  and statistical thermal models}, Phys. Rev. C \textbf{109}, 014907 (2024),
  \texttt{2302.13852}. \doiwoc{10.1103/PhysRevC.109.014907}

\bibitem{Du:2023efk}
L.~Du, {Bulk medium properties of heavy-ion collisions from the beam energy
  scan with a multistage hydrodynamic model}, Phys. Rev. C \textbf{110}, 014904
  (2024), \texttt{2401.00596}. \doiwoc{10.1103/PhysRevC.110.014904}

\bibitem{Du:2024pbd}
L.~Du, {Multimessenger study of baryon-charged QCD matter in heavy-ion
  collisions}, Phys. Lett. B \textbf{861}, 139270 (2025), \texttt{2408.08501}.
  \doiwoc{10.1016/j.physletb.2025.139270}

\bibitem{JETSCAPE:2020mzn}
D.~Everett et~al. (JETSCAPE), {Multisystem Bayesian constraints on the
  transport coefficients of QCD matter}, Phys. Rev. C \textbf{103}, 054904
  (2021), \texttt{2011.01430}. \doiwoc{10.1103/PhysRevC.103.054904}

\bibitem{JETSCAPE:2024cqe}
R.~Ehlers et~al. (JETSCAPE), {Bayesian Inference analysis of jet quenching
  using inclusive jet and hadron suppression measurements} (2024),
  \texttt{2408.08247}.

\end{thebibliography}
\end{document}